# Multi-objective CFD optimization of an intermediate diffuser stage for PediaFlow pediatric ventricular assist device

Mansur Zhussupbekov[1], JingChun Wu[2], Greg W Burgreen[3], Scott Stelick[1], Jeongho Kim[4], and James F Antaki[1]*

[1] Meinig School of Biomedical Engineering, Cornell University, Ithaca, NY, USA
[2] Advanced Design Optimization, LLC, Irvine, CA, USA
[3] Center for Advanced Vehicular Systems, Mississippi State University, Starkville, MS
[4] Department of Biomedical Engineering, Daejeon Institute of Science and Technology, Daejeon, Korea
* Correspondence: antaki@cornell.edu

**Abstract**

**Background:** Computational fluid dynamics (CFD) has become an essential design tool for ventricular assist devices (VADs), where the goal of maximizing performance often conflicts with biocompatibility. This tradeoff becomes even more pronounced in pediatric applications due to the stringent size constraints imposed by the smaller patient population. This study presents an automated CFD-driven shape optimization of a new intermediate diffuser stage for the PediaFlow pediatric VAD, positioned immediately downstream of the impeller to improve pressure recovery.

**Methods:** We adopted a multi-objective optimization approach to maximize pressure recovery while minimizing hemolysis. The proposed diffuser stage was isolated from the rest of the flow domain, enabling efficient evaluation of over 450 design variants using Sobol sequence, which yielded a Pareto front of non-dominated solutions. The selected best candidate was further refined using local T-search algorithm. We then incorporated the optimized front diffuser into the full pump for CFD verification and in vitro validation.

**Results:** We identified critical dependencies where longer blades increased pressure recovery but also hemolysis, while the wrap angle showed a strong parabolic relationship with pressure recovery but a monotonic relationship with hemolysis. Counterintuitively, configurations with fewer blades (2-3) consistently outperformed those with more blades (4-5) in both metrics. The optimized two-blade design enabled operation at lower pump speeds (14,000 vs 16,000 RPM), improving hydraulic efficiency from 26.3% to 32.5% and reducing hemolysis by 31%.

**Conclusion:** This approach demonstrates that multi-objective CFD optimization can systematically explore complex design spaces while balancing competing priorities of performance and hemocompatibility for pediatric VADs.

# Introduction

Computational fluid dynamics (CFD) has become an invaluable design tool for blood-wetted medical devices including rotodynamic blood pumps such as left ventricular assist devices (VADs). VADs are unique from traditional turbomachinery by virtue of their requirement for miniaturization and biocompatibility. CFD enables detailed evaluation of flow path features that are not readily prescribed by traditional textbook formulae and empirical guidelines, such as hub and shroud profiles, blade leading- and trailing-edge angles, blade wrap and length, number of blades, and volute dimensions.[1,2] The adoption of magnetically-levitated (maglev) rotors further distinguishes VADs from traditional pumps, requiring non-standard shapes of rotor hubs that house magnets and large secondary flow paths to improve washing. These geometric parameters are complex, interrelated, and affect both performance and hemocompatibility. While the inverse design method can compute aspects of the flow path based on a specified pressure distribution, CFD evaluation is used to further optimize these design choices.[3,4]

Three distinct approaches to CFD-based design have emerged and continue to be used concurrently by different groups, summarized in **Supplementary Table S1**. The simplest approach involves manual comparison of a limited set of design variants to select superior geometries.[5–10] While straightforward, this method often relies on intuition and trial-and-error exploration. A more systematic approach involves methodically varying individual design parameters to reveal correlations between geometric features and performance metrics, as well as to understand the interplay between competing objectives like efficiency and biocompatibility.[11–13] However, both manual approaches are labor intensive and limited by time, resources, and the experience of the designer to manage and tradeoff a multitude of parameters. Consequently, the end design is unlikely to be *optimal* in the formal sense of the word. The most computationally intensive approach, automated CFD-driven optimization, addresses these limitations by enabling exploration of vast design spaces with minimal human intervention, systematically determining optimal combinations of design parameters that maximize or minimize specified objective function(s) while satisfying geometric and performance constraints.[14–20]

The present work concerns the ongoing evolution of the PediaFlow VAD – a pediatric, fully maglev, rotodynamic blood pump intended to provide chronic circulatory support for children with congenital or acquired heart failure.[21] Rigorous mathematical modeling and optimization have been at the core of PediaFlow's development since its conception, yielding a miniature design comparable to AA-battery in size.[11,22] For the fifth-generation device (PF5), shown in **Figure 1a**, Wu et al. (2022) increased the flow capacity to 4.0 L/min and improved efficiency of the previous-generation pump by increasing the inflow and outflow diameters and further optimizing the blood flow path using the inverse design method and CFD.[23]

**Figure 1b** shows the static pressure in PF5 at 1.5 LPM and 16,000 RPM, plotted along the axial position. The flow path of the PediaFlow consists of a mixed-flow impeller with 4 blades in a conical inlet, a 1.5-mm annular fluid gap region, and a 3-vane diffuser in a conical aft-housing. Kinetic energy imparted by the impeller develops dynamic head which is then recovered (partially) as static pressure in the diffuser. Achieving sufficient magnetic stiffness of the maglev rotor necessitated increasing the rotor hub length, resulting in an unconventional design with a long axial separation between the impeller and diffuser stages. However, this extended annular gap presented a hemodynamic challenge: viscous energy losses arose due to high circumferential velocity of blood, further exacerbated by Taylor vortices between the cylindrical hub and shroud surfaces.[24,25]

To address these losses, we evaluated adding a second set of diffuser blades (*front diffuser*) located immediately downstream of the impeller to mitigate high velocity swirling flow and convert dynamic head into static pressure more efficiently (see **Figure 1b**). Our hypothesis was that this unorthodox design modification would improve pump efficiency by achieving the desired operating point at a lower rotor speed.

When improving VAD performance, any gains in hydraulic characteristics must be carefully weighed against potential impact on hemocompatibility. This inherent trade-off between performance and blood damage makes this task particularly well-suited for multi-objective optimization. Leveraging automated CFD-driven shape optimization, we sought to determine the optimal number of blades and their ideal shape to maximize pressure recovery of the new front diffuser stage while minimizing hemolysis.

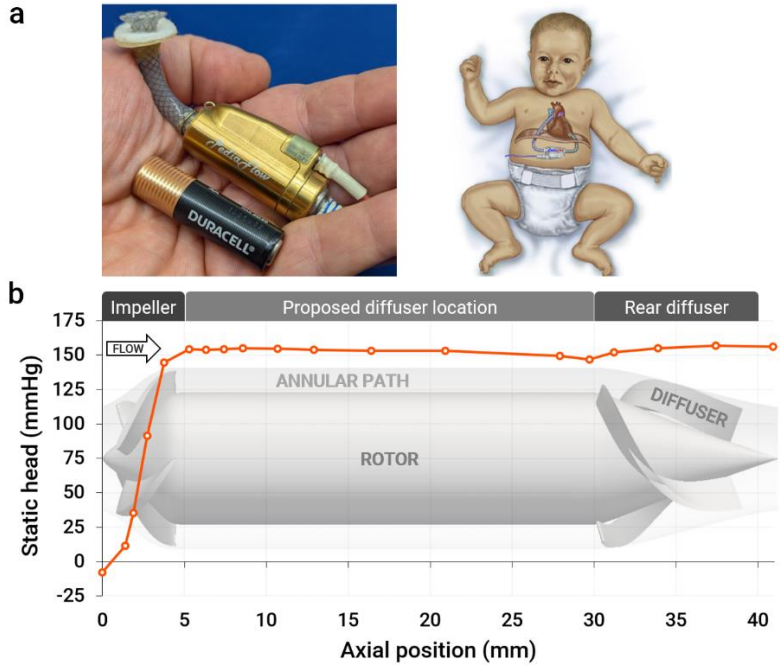

**Figure 1.** (a) The fifth-generation PediaFlow VAD compared to an AA-battery. (b) Static head produced by PF5 operating at 1.5 LPM and 16,000 RPM, plotted along the axial position. The image of the flow path is overlaid on the plot, and the labels above the plot indicate the location of the impeller, the proposed front diffuser, and the existing rear diffuser.

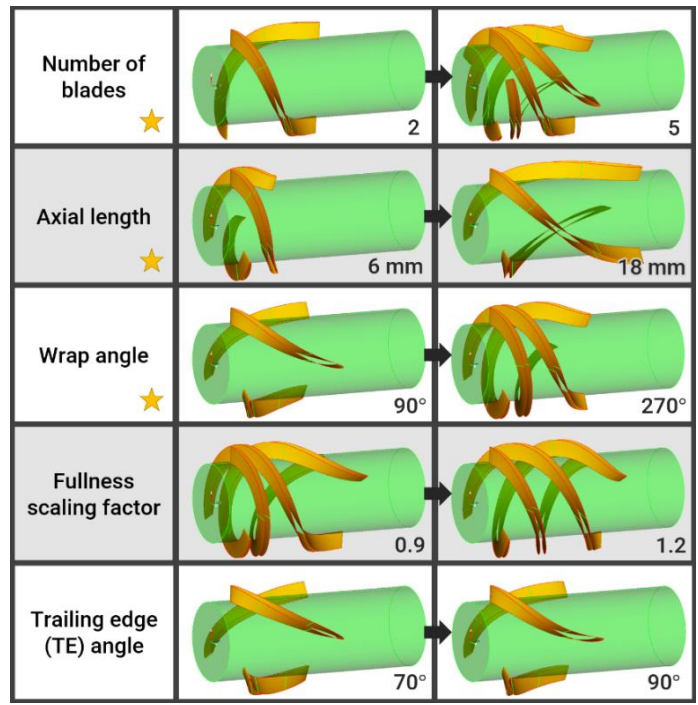

**Figure 2.** Input design parameters and their limits considered in the exploration study of an isolated diffuser stage. Parameters that showed significant influence on pressure recovery are marked with a star. The stationary diffuser blades are highlighted in yellow, and the rotor surface (hub) is shown for reference in green.

## Materials and Methods

### Shape parametrization

The straight section of the annular passage with the proposed diffuser stage was isolated from the rest of the flow path as depicted in **Figure 2**. The blade geometry was fully parametrized using a B-spline in the CAESES® computer-aided drawing software (Friendship Systems AG, Potsdam, Germany). The number of blades was constrained to vary from 2 to 5, and the blade shape was defined by four variables: axial length, wrap angle, the trailing edge (TE) angle, and the scaled area under the camber curve, referred to as the *fullness scaling factor*.

Independent variation of axial length and wrap angle could lead to designs with extreme ratios between the two parameters. In preliminary testing, it was found that excessively low ratios resulted in diffuser designs not providing adequate flow guidance, leading to undesirable flow separation. Conversely, designs with excessively high ratios were found to be overly congested, causing flow blockage. Designs with these extreme ratios failed to generate pressure gain. To address this issue, the axial length of the diffuser blades was directly specified, while the blade wrap angle was controlled via a *wrap angle to axial length ratio* (degree/mm), constrained to prevent the generation of unproductive extreme designs. Constraints for this and the remaining input variables are detailed in **Supplementary Table S2**.

The radial gap between the tip of the diffuser blades and the rotor surface was fixed at 0.1 mm. The leading edge (LE) of the blades was placed 2 mm from the inlet of the isolated annular gap domain. The blade LE angle, defined as the inclination of the camber line of the blade in the meridional plane with respect to the plane perpendicular to the axis of rotation, was fixed at $\alpha_1 = \tan^{-1}\left(\frac{|\bar{u}_{mer}|}{|\bar{u}_{circ}|}\right) + \varepsilon = 7.26° + 3° = 10.26°$, where $\bar{u}_{mer}$ and $\bar{u}_{circ}$ were the average meridional and circumferential velocities specified at the inlet boundary and $\varepsilon$ is a small angle added to account for the expelling effect of the blades. Although impeller outflow is typically nonuniform with velocity variations from hub to shroud, we simplified the optimization by using a fixed average LE angle for the front diffuser. Similarly, we adopted a fixed TE angle, making the blade profile constant across the entire span.

### Automated mesh generation

The CAESES module was coupled with OpenFOAM for meshing and CFD simulation.[26] STL files created by the CAESES were imported into OpenFOAM's *snappyHexMesh* meshing utility that generated a 3-dimensional mesh composed of predominantly hexahedral cells. Increased levels of mesh refinement were applied at the diffuser blade surface, along the leading and trailing edges, and the blade tip gap. To ensure robust execution of the parametric CAD generation and meshing scripts, the entire domain was modeled and meshed instead of a periodic segment. Mesh counts ranged from 2.2 million cells for the shortest 2-blade diffuser to 5.9 million cells for the longest 5-blade diffuser.

### CFD boundary conditions

The inlet velocity boundary condition (BC) for the isolated annular gap domain was estimated by sampling the velocity field from a simulation of the full PF5 pump operating at 1.5 L/min, 14,000 RPM. The area-averaged values of the meridional and circumferential velocity components were computed on a cross-section 1 mm downstream from the impeller stage and were prescribed as a uniform velocity profile at the inlet of the isolated domain. Similarly, sampled area-averaged values of the turbulence kinetic energy, specific dissipation rate, and kinematic eddy viscosity in the *k-ω* SST turbulence model were prescribed at the inlet. A rotating wall velocity of 14,000 RPM was prescribed at the hub surface, and no-slip BCs were applied at the diffuser blade and shroud surfaces. Wall functions were used for the turbulence quantities, and the y+ < 1 mesh criterion was satisfied on all surfaces. Blood was treated as an incompressible Newtonian fluid with a density of 1060 kg m$^{-3}$ and viscosity of 3.5 cP. The assumption of Newtonian fluid was applicable because the shear rate levels in the computational domain far exceeded 100 s$^{-1}$. (See in **Supplementary Table S3**.) After reaching a converged solution, the pressure gain (ΔP) was calculated as the difference between the outlet pressure (zero) and the average pressure on the inlet boundary.

*Hemolysis computation*

Hemolysis was computed using the Giersiepen-Wurzinger power law model[27,28] employing the Eulerian approach proposed by Garon & Farinas.[29] Specifically, the asymptotically consistent formulation described in Farinas et al.[30] was implemented in OpenFOAM. The constants published by Heuser & Opitz[31,32] were used as they are applicable across a broader range of shear stress values and demonstrated an excellent correlation coefficient in the study by Taskin et al.[33] The scalar shear stress in the hemolysis source term was computed following Faghih & Sharp.[34] When using a uniform velocity BC at the inlet and no slip BC at the walls, high shear stresses may arise in the cells adjacent to the inlet and wall boundaries. Therefore, the hemolysis source term was disabled in the first 1 mm of the domain to avoid artificial hemolysis.

Since most numerical models of hemolysis do not predict an accurate absolute value,[35] the relative hemolysis was evaluated using the Relative Hemolysis Index (RHI), similar to the approach in Gil et al.,[36] and calculated as RHI = HI/HI$_{nominal}$, where HI$_{nominal}$ represents the hemolysis generated by the annular section of the flow path with no diffuser blades at the same flow rate and rotor speed.

*Mesh independence study*

To assess mesh independence, a representative design featuring 2 blades was selected, and the pressure gain and hemolysis were computed at three levels of mesh refinement. The target difference in computed values between successive levels of refinement was 1%. The background grid was refined progressively by a factor of 1.15 in every dimension, while the *snappyHexMesh* settings remained unchanged as mesh manipulations in this stage were performed relative to the background grid. This yielded three mesh versions: coarse, with 2.4 million cells; medium, with 3.4 million cells; and fine, with 4.7 million cells. The discrepancy in pressure rise between the medium and fine mesh versions was 0.35%, and the hemolysis values were within 0.86%. Consequently, the medium level of mesh refinement was adopted for the optimization process.

The OpenFOAM module, encompassing mesh generation, flow solution, and hemolysis evaluation, was automated using a Bash script to output ΔP and RHI values that were fed back to CAESES.

*Optimization Procedure*

The optimization problem addressed two competing objectives: maximizing pressure gain (ΔP) generated by the diffuser while minimizing hemolysis. However, following standard convention for computational optimization, the pressure objective was formulated as minimizing -ΔP. We implemented a two-stage optimization strategy, illustrated in **Figure 3**: an exploration stage using the Sobol sequence, followed by a local optimization using a gradient-free T-search algorithm. In general, the Sobol stage reveals trends in the design space and identifies best candidate(s), while the T-search algorithm further probes the vicinity of a selected design point with the goal of converging closer to a true local optimum.[37]

We performed two rounds of Sobol exploration. Exploration 1 focused solely on the objective function of pressure recovery (-ΔP) and probed five design variables: number of blades, axial length, wrap angle, the TE angle, and the fullness scaling factor (**Figure 2**). As a general rule,[38] number of repeated analyses is proportional to the third power of the number of design variables, $n^3$. However, since the number of blades was an integer parameter varying from 2 to 5, we evaluated sizeof($n_{blades}$) ×$n_{shape\_variables}^3$ = 4×4³ = 256 design variants, which constitutes a more comprehensive search than $5^3$ variants.

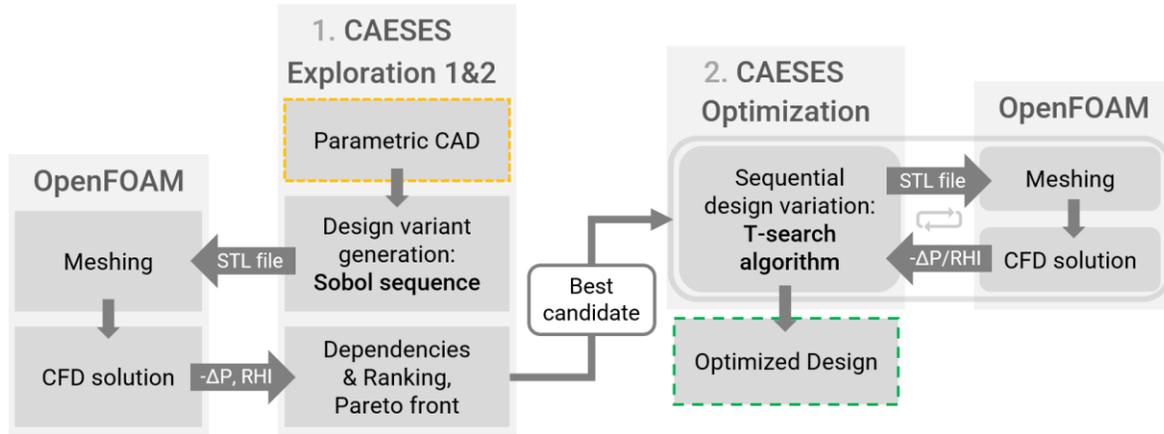

**Figure 3.** Schematic depiction of the optimization process and communication between the CAESES and OpenFOAM modules.

For Exploration 2, both pressure recovery and hemolysis were included as distinct objective functions. The input variables that demonstrated weak sensitivity in the previous stage were eliminated, and constraints for the remaining parameters were adjusted to more favorable ranges. 64 design variants were generated for each specified number of blades to accommodate custom parameter ranges. The best candidate identified in this round underwent further optimization using the T-search algorithm.

All statistical analyses were conducted using Python with SciPy and scikit-learn libraries, and the details of the methodology are provided in **SUPPORTING INFORMATION**.

The study was conducted using a single multi-core processor: Intel Xeon Gold 2.4GHz, 3.7GHz Turbo, 20 cores with 2 threads per core. The initial exploration study, comprising 256 design variants, was completed in 16 days. The second exploration study, involving evaluations for both pressure and hemolysis, encompassed 196 design variants and was completed in 10 days.

*CFD verification and in vitro validation*

To verify the hydraulic performance of the optimized front diffuser, it was introduced into the flow path of the full pump, positioned 2 mm downstream from the trailing edge of the impeller blades. The CFD verification was performed by Advanced Design Optimization (ADO), LLC (Irvine, CA, USA). A multiblock structured hexahedral mesh with boundary orthogonality was generated in ADO mesh using elliptical method. CFD evaluation was performed in ANSYS-CFX using frozen-rotor approach and SST turbulence model, treating blood as Newtonian fluid. Hemolysis was computed using the Eulerian method with model constants by Fraser et al.[39] A mesh independence study with 1% convergence criterion resulted in a final mesh containing 4.6 million cells, shown in **Supplementary Figure S1**. The modified pump was simulated at 14,000 and 16,000 RPM with a flow rate of 1.5 L/min, and the resulting pressure head, efficiency, and hemolysis were compared to those of the original pump without the front diffuser stage.

For *in vitro* validation, the optimized front diffuser design was fabricated using rapid prototyping on a stereolithography (SLA) printer with a 25-μm resolution. A 0.3-mm thick cylindrical shroud was added to support the blades (similar to **Figure 7a**), which reduced the blade height by 20% from the simulated design. This diffuser stage was integrated into the PF5 pump prototype, and HQ curves were obtained using a 35% glycerol solution in a benchtop flow loop.

## Results

*Exploration 1*

The first round of exploration included 256 design variants across a broad parameter space to identify the most influential design variables. 46 designs failed to produce pressure gain and were excluded from the analysis. Statistical analysis of the resulting design space revealed substantial variance (coefficient of variation, CV = 41.4%), reflecting the complex, interdependent relationships between parameters. (See **Supplementary Table S4**.)

**Figure 4** shows pressure recovery, -ΔP, plotted against the five design variables. Number of blades demonstrated a clear categorical effect ($R^2 = 0.10$, $p < 0.001$), with 2-blade designs achieving pressure gains up to 40 mmHg while 5-blade configuration failed to exceed 30 mmHg. Axial length showed a significant monotonic relationship with pressure recovery ($R^2 = 0.08$, $p < 0.001$), with longer diffuser blades producing greater pressure gains. The wrap angle to axial length ratio emerged as the strongest predictor of pressure recovery, demonstrating a distinct parabolic relationship and accounting for 35% of pressure variance ($R^2 = 0.35$, $p < 0.001$). As seen from **Figure 4c,** optimal pressure recovery occurred at intermediate values of wrap angle to axial length ratio, at 10-20 degree/mm. In contrast, TE angle and fullness scaling factor demonstrated no significant correlation with pressure recovery ($R^2 = 0.00$, $p = 0.323$ and $p = 0.326$, respectively).

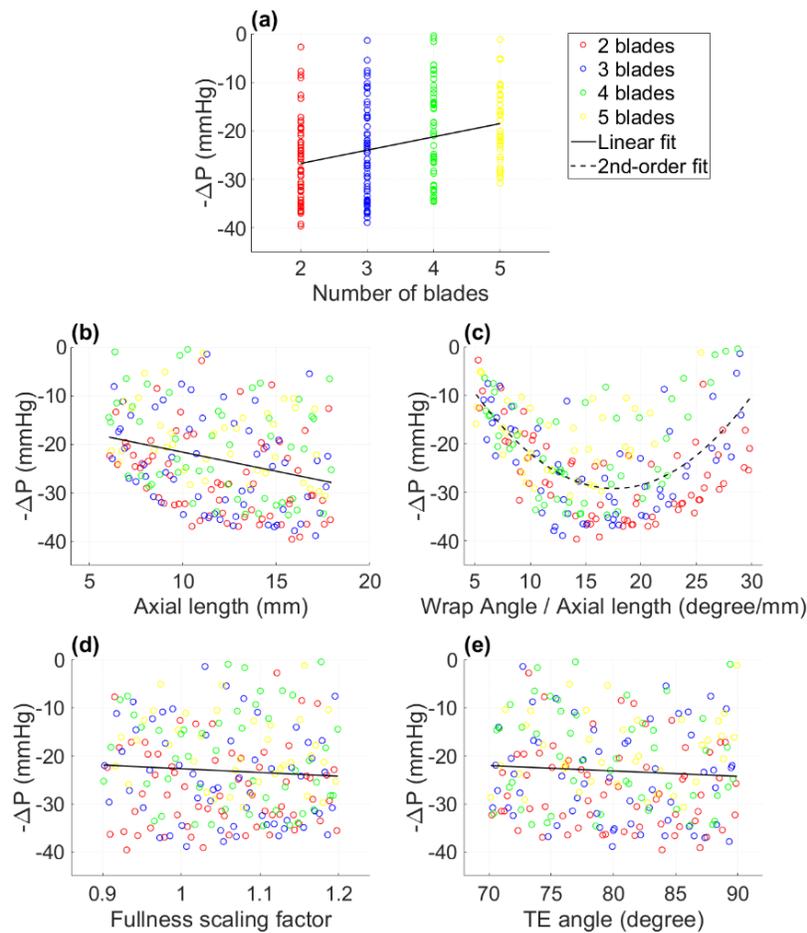

**Figure 4.** Results of the first exploration study showing pressure recovery, -ΔP, plotted against input variables.

*Exploration 2*

Based on these initial findings, the second exploration stage focused on a narrower design space: (1) 5-blade configuration was excluded, (2) statistically insignificant parameters, TE angle and fullness scaling factor, were set to a fixed value, and (3) 2-, 3-, and 4-blade configurations were explored separately (64 design variants each), with constraints for axial length and wrap angle to axial length ratio adjusted to custom ranges for each blade number. (See **Supplementary Table S2**.) Finally, Exploration 2 involved two objective functions: maximizing ΔP (minimizing -ΔP) and minimizing hemolysis (RHI).

The second exploration demonstrated substantially stronger parameter relationships and reduced design space variance. CV decreased to 11.6% for 2-blade designs, 18.8% for 3-blade designs, and 17.8% for 4-blade designs (**Supplementary Table S5**). Consistent with prior observations, the 2-blade designs consistently outperformed configurations with more blades, achieving mean pressure recovery of 34.3 mmHg, compared to 30.7 mmHg for 3 blades and 28.2 mmHg for 4 blades. Cross-blade statistical analysis revealed a significant linear trend ($F = 24.4$, $p < 0.001$, $\eta^2 = 0.205$), where each additional blade reduced the mean pressure recovery by 3.06 mmHg. (**Supplementary Table S6**).

Blade number showed an even stronger effect on hemocompatibility ($F = 62.2$, $p < 0.001$, $\eta^2 = 0.397$), with each additional blade increasing hemolysis by 0.11 RHI units. 2-blade designs achieved the lowest hemolysis (mean RHI = 1.66), compared to 3-blade (RHI = 1.78) and 4-blade designs (RHI = 1.88).

**Figure 5** presents the parameter relationships for 2-blade configuration. Axial length demonstrated strong linear relationships with both objectives: longer blades improved pressure recovery ($R^2 = 0.32$, $p < 0.001$) but also increased hemolysis ($R^2 = 0.52$, $p < 0.001$). The wrap angle to axial length ratio showed a strong parabolic relationship with pressure recovery ($R^2 = 0.28$, $p < 0.001$) and a negative linear relationship with hemolysis ($R^2 = 0.34$, $p < 0.001$). Notably, axial length emerged as the strongest predictor of hemolysis across all blade number configurations ($R^2 = 0.52$-$0.76$), while wrap angle to axial length ratio was the primary driver of pressure recovery optimization ($R^2 = 0.28$-$0.47$). (See **Supplementary Table S7** and **Supplementary Figure S2** and **Supplementary Figure S3**.)

*Selecting the best candidate*

**Figure 6a** shows the plot of the pressure rise against hemolysis for 2-blade configuration, highlighting the inherent trade-off between these two objectives. The non-inferior envelope, known as the Pareto front, can be discerned along the lower edge of the point cloud, highlighted in yellow. This delineates the optimal compromise between achieving the lowest hemolysis for a given pressure recovery. All designs situated along the Pareto front are deemed potential optimal candidates for selection, and the selection of the most suitable candidate typically involves a matter of preference or prioritization of one objective function over the other.

The primary goal of adding this additional front stator stage was to enable operation at a lower rotor speed (14,000 RPM vs 16,000 RPM) while maintaining the same operating point (flow rate and pressure). At the selected flow rate of 1.5 L/min, the baseline PF5 pump produced 156 mmHg at 16,000 RPM and 117 mmHg at 14,000 RPM. Therefore, the added stator stage needed to generate 39 mmHg of additional pressure head to compensate for the speed reduction. In Figure 6a, the design point highlighted in red achieved precisely this target, producing 39.2 mmHg. When ranking the design variants using the ratio of -ΔP to RHI (**Figure 6b** and **6c**), this same design emerged as the best candidate, confirming its balanced hydraulic and hemolytic performance.

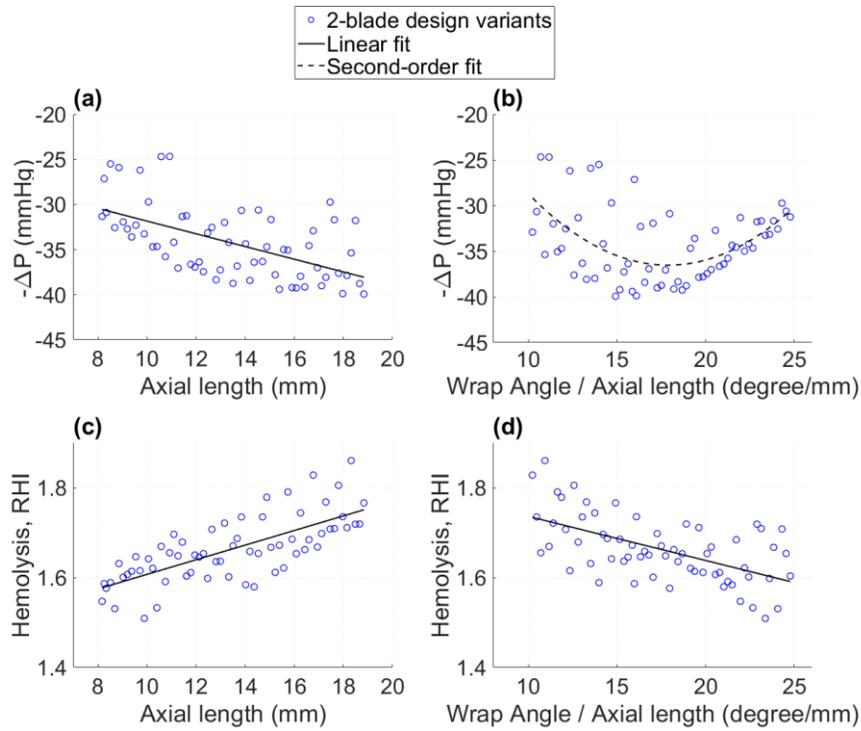

**Figure 5.** Results of Exploration 2 for the 2-blade configuration. (a)-(b) Pressure recovery, -ΔP, plotted against input variables. (c)-(d) Hemolysis, expressed as RHI, plotted against the input variables.

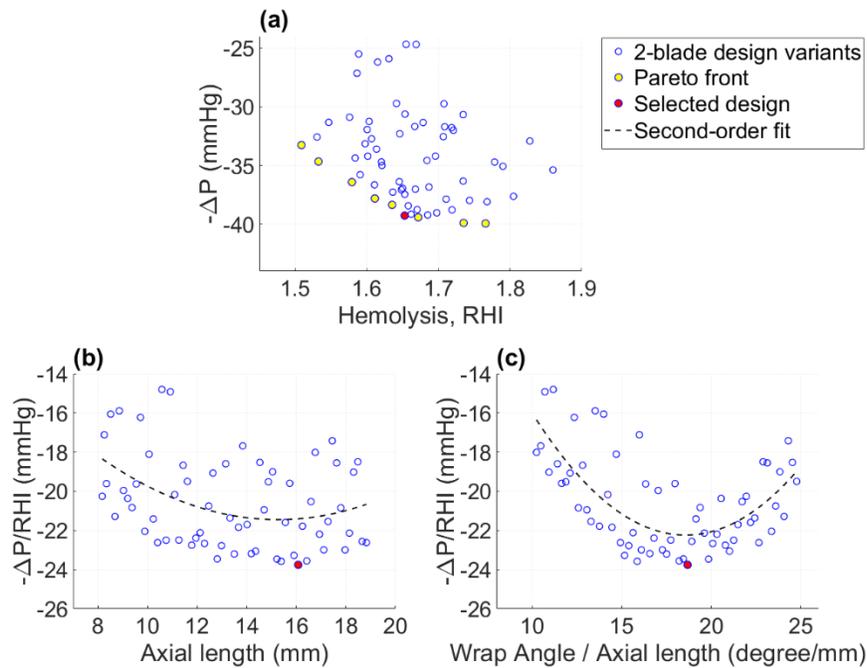

**Figure 6.** Selection of the best candidate based on the trade-off between the two objective functions. (a) Plotting the pressure recovery against the hemolysis reveals the Pareto front, highlighted in yellow, along the lower bound of the scatter plot. (b)-(c) The utility function, combining the two objectives into the ratio of -ΔP to RHI, plotted against the input variables. The best candidate is indicated in red.

*Local optimization*

For the final stage of optimization, the selected candidate was further optimized using a T-search algorithm with the objective of improving the ratio -ΔP/RHI. **Supplementary Figure S4** depicts the evolution of this combined objective function from the initial design to the optimized design. The T-search optimization improved the ratio -ΔP/RHI from 23.76 to 24.0, primarily attributed to the reduction in hemolysis from RHI 1.65 to 1.63.

The resulting optimized design is showcased in **Figure 7a**. The velocity vectors plotted at mid-span of the blades, shown in **Figure 7b**, reveal well-attached flow patterns. **Figure 7c** illustrates the Hemolysis Index (HI) on the hub, shroud, and blade surfaces. Highest shear stress levels occurred at the leading edge of the blades and at the blade tips. Flow emanating from the blade tip gap impinges on the pressure side of the blades, leading to the formation of a ribbon-like trail of elevated HI on the blade surface.

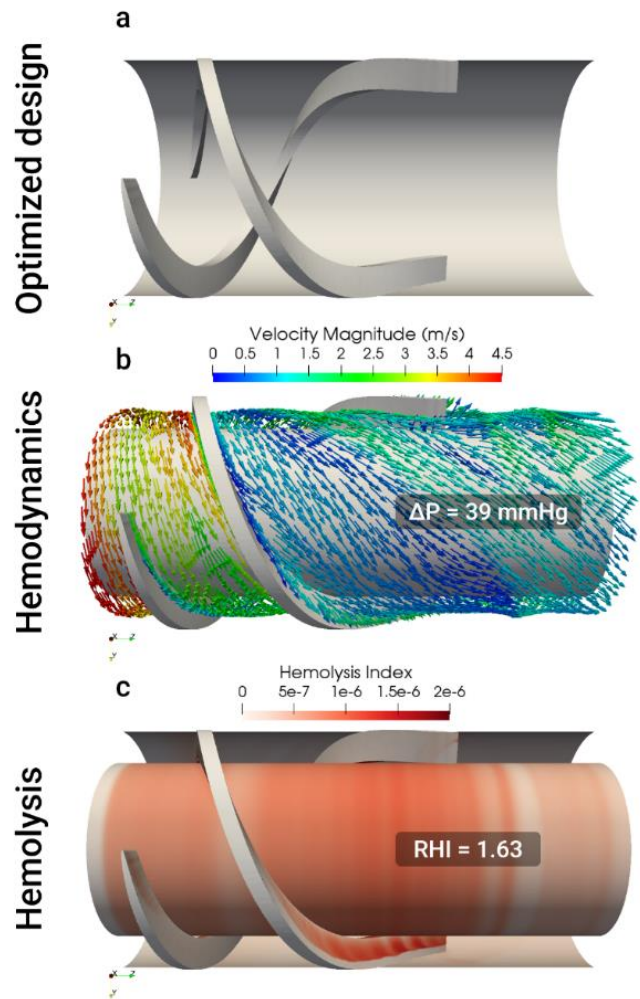

**Figure 7.** The optimized front diffuser design, featuring 2 blades, achieved a pressure rise of 39 mmHg in an isolated domain. Hemolysis within this section of the flow path increased by a factor of 1.63 compared to the reference (no-blade) configuration. (a) The diffuser blades shown attached to the shroud surface. (b) Visualization of velocity vectors at mid-span, with flow progressing from left to right. (c) Hemolysis Index shown on blood contacting surfaces.

*CFD verification and in vitro validation*

**Figure 8** shows the computed pressure head plotted along the axial position for the PF5 pump with the optimized 2-blade front diffuser compared against the baseline pump. The addition of the front diffuser enabled achieving the same operating point (160±5 mmHg at 1.5 L/min) at a reduced speed of 14,000 RPM versus 16,000 RPM for the baseline. At these comparable conditions, the modified design showed higher hydraulic efficiency (32.5% vs 26.3%) and lower power consumption (1.70W vs 1.97W), as detailed in **Table 1**. Importantly, the computed NIH decreased by 31% from 0.0192 to 0.0132 g/100L, suggesting improved hemocompatibility.

**Figure 9** shows the *in vitro* HQ curves for the pump fitted with 3D-printed front stator stage, revealing hydraulic performance across a wider range of flow rates. At 1.5 L/min, *in vitro* pressure head was 154 mmHg at 14,000 RPM and 210 mmHg at 16,000 RPM – 6.7% and 9.5% lower than CFD-predicted values, respectively, likely due to the reduction in blade height necessitated by rapid prototyping.

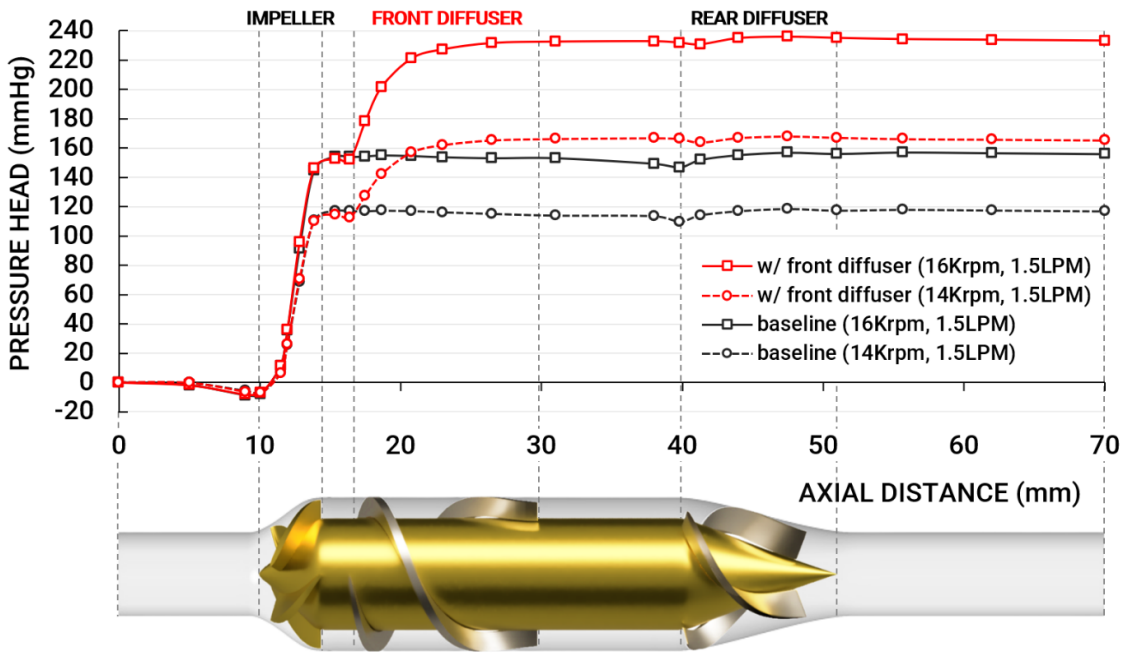

**Figure 8.** CFD performance of the optimized front diffuser in the PF5 pump at 1.5 LPM and speed of 14,000 and 16,000 RPM compared to the original pump (baseline). The image of PF5 with added front diffuser is overlaid on the plot to match the axial position; original PF5 not shown.

**Table 1.** CFD pump performance at 1.5 L/min with and without the added front diffuser. NIH was computed using the power-law model with coefficients by Fraser et al.[39]

| Pump version | Speed (RPM) | ΔP (mmHg) | Power (W) | Efficiency (%) | Computed NIH (g/100L) |
|---|---|---|---|---|---|
| PF5 baseline | 14,000 | 117 | 1.44 | 27.1 | 0.0119 |
|  | 16,000 | 156 | 1.97 | 26.3 | 0.0192 |
| PF5 w/ front diffuser | 14,000 | 165 | 1.70 | 32.5 | 0.0132 |
|  | 16,000 | 232 | 2.45 | 31.5 | 0.0222 |

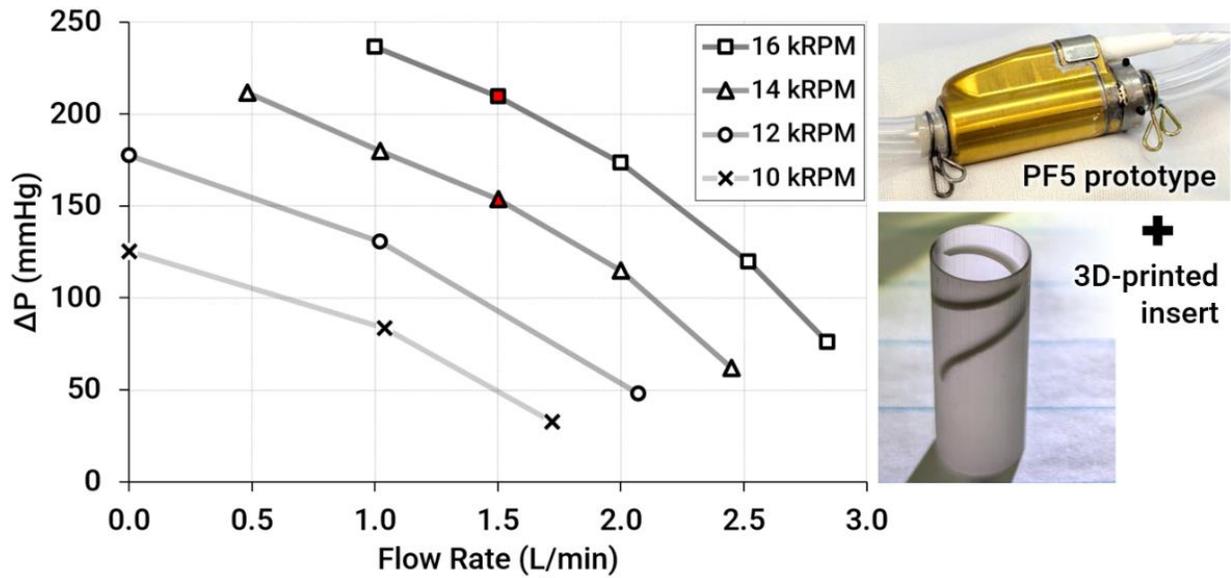

**Figure 9.** *In vitro* HQ curves for the pump fitted with 3D-printed front stator stage. Markers filled in red indicate the flow rate at which the optimization was performed.

**Discussion**

In this study, we sought to improve the performance of the PediaFlow pediatric VAD by introducing an additional diffuser stage in the long annular space between the impeller and existing diffuser stages. We employed multi-objective optimization to design the diffuser blades, where the two concurrent objectives were maximizing pressure recovery and minimizing hemolysis. The addition of the optimized diffuser blades enabled achieving the same operating point at a lower pump speed, resulting in improved hydraulic efficiency and reduced hemolysis.

The relationship between design parameters and performance metrics showed complex, often competing trends that revealed distinct design priorities for each objective. Across all blade configurations, axial length emerged as the dominant predictor of hemolysis ($R^2 = 0.52$-$0.76$), although the effect on pressure recovery was still significant ($R^2 = 0.15$-$0.32$). Conversely, the wrap angle to axial length ratio exhibited a strong parabolic relationship with pressure recovery ($R^2 = 0.28$-$0.47$) but a less pronounced monotonic relationship with hemolysis ($R^2 = 0.10$-$0.34$). These competing objectives created a Pareto front, seen in **Figure 6a**, providing a rational basis for optimal design selection. These interdependencies echo findings from previous studies by Wiegmann et al. (2018)[12] and Li et al. (2022)[13], though our work provides more comprehensive characterization of these relationships for diffuser blades specifically. The Pareto front observed in our study resembles those reported by Zhu et al. (2010)[15] for hydraulic performance of diffuser blades and Ghadimi et al. (2019)[18], who optimized for efficiency and hemolysis in a centrifugal pump. In a recent state-of-the-art study, Escher et al. (2024)[20] further expanded this methodology by integrating both hydraulic and electromagnetic optimization domains, revealing a comprehensive Pareto front that navigates the trade-offs between pump dimensions, motor performance, and hemocompatibility for a cavopulmonary assist device.

The best candidate among Pareto-optimal solutions was selected based on achieving the target pressure recovery (39 mmHg) needed to maintain pump performance at a reduced speed of 14,000 RPM (down from 16,000 RPM). A particularly encouraging finding was that this same design emerged as optimal when ranked by the ratio of pressure recovery to hemolysis (-ΔP/RHI). The second-order polynomial fits, shown as dashed lines in **Figure 6b** and **6c** (axial length $R^2 = 0.15$, $p = 0.004$, and wrap angle to axial length ratio $R^2 = 0.46$, $p < 0.001$), indicate the existence of optimal ranges for both input parameters with respect to this combined performance metric, with the selected design situated at the global optimum. The local optimization using the T-search algorithm yielded only modest improvements in -ΔP/RHI, suggesting that the exploration stage had already identified a near-optimal design.

When implemented in the full pump simulation, the modified design successfully achieved the same operating point (160±5 mmHg at 1.5 L/min) at lower speed while improving efficiency from 26.3% to 32.5% and reducing computed hemolysis by 31%. While we expected improvement in hydraulic efficiency (reduction in losses) through better energy conversion, the net effect on hemolysis was not self-evident as the addition of new blade surfaces would introduce additional high-shear regions at leading edges and within blade tip gaps. Escher et al. (2022)[40] showed that hydraulic energy dissipation and hemolysis are strongly correlated. **Supplementary Figure S5** confirms that the modified pump achieved both reduced NIH and energy dissipation at equivalent operating points, which aligns with Escher's findings and suggests that the hydraulic benefits outweigh localized shear effects.

Our exploration revealed several counterintuitive findings. Most notably, configurations with fewer blades consistently outperformed those with more blades in both pressure recovery and hemolysis metrics. This contradicts traditional turbomachinery recommendations that suggest using more blades for better flow guidance (as the number of blades increases, the Busemann slip factor tends to unity).[2] However, those recommendations were developed for large-capacity industrial pumps, in which inertial forces are far more dominant.[6] Additionally, ΔP showed minimal dependence on the TE angle. This weak correlation may be explained by the rotor imparting circumferential momentum to the fluid through viscous forces in the annular gap, effectively diminishing the benefit of completely removing swirling flow at the front diffuser's trailing edge. It should be noted that we are reporting empirical findings from our specific application rather than proposing universal design recommendations. These counterintuitive relationships merit further investigation across broader operating conditions and geometric configurations.

Several limitations of this study should be acknowledged. The optimization focused on steady flow at a single operating point, whereas clinical implementation would require acceptable performance across a range of conditions and pulsatile flow. Additionally, while hemolysis was considered in the optimization, other aspects of hemocompatibility such as thrombosis risk require further investigation. Finally, manufacturability constraints were not included in the optimization: the large blade wrap angle may pose fabrication challenges at this miniature scale.

**Conclusion**

This work demonstrates that CFD-driven optimization provides valuable insights when addressing open-ended design problems within constrained spaces while balancing competing performance objectives. The systematic progression from broad (weak correlations, high variance) to focused exploration (strong correlations, low variance) demonstrated the effectiveness of statistical analysis in guiding design space refinement for complex, multi-objective problems. Future work should extend this methodology to multi-point optimization, incorporating broader operating conditions and additional hemocompatibility metrics.

**Acknowledgments**

This work was supported by the National Institutes of Health grant R01HL089456 and the U.S. Army Medical Research Acquisition Activity Project Number W81XWH2010387. We wish to express our gratitude to Friendship Systems AG for providing a free academic license for CAESES and their technical support.

# SUPPORTING INFORMATION

**Supplementary Table S1.** Summary of the CFD-based design approaches to VAD optimization with notable examples.

| Method | Publication | Optimization method | Pump flow type, bearing system | Design parameters | Performance metric(s) or objective function(s) | Improvement/findings | *In vitro* validation |
|---|---|---|---|---|---|---|---|
| Manual comparison of design variants | Arvand et al. (2004)[5] | Manual comparison of 3 design variants | Mixed flow, blood-immersed mechanical bearing | Impeller design variants | Hydraulic performance (HQ), hemolysis | Identified the design variant with superior hydraulic performance and lowest hemolysis | Yes: hydraulic performance, hemolysis |
| | Untaroiu et al. (2005)[6] | Manual comparison of 3- and 6-bladed diffusers. The blade geometry was designed using conventional formulae and CFD refinement | Axial flow, maglev | Number of diffuser blades | Pressure head, efficiency, axial force exerted on the maglev rotor | 6-bladed diffuser showed superior hydraulic performance, but the 3-bladed version matched it with a 500 RPM rise in rotor speed and offered better manufacturability | Yes: hydraulic performance |
| | Wu et al. (2005)[7] | Manual comparison of the initial design and 2 variations of secondary blade design | Centrifugal, maglev | Back clearance gap, addition of secondary blades and number thereof, TE angle | Computed net leakage flow through the back clearance gap | Net antegrade flow with minimal zones of retrograde flow was achieved | Yes: flow visualization |
| | Zhang et al. (2007)[8] | Manual comparison of 3 design variants | Centrifugal, maglev | Impeller design variants | Pressure head, hemolysis | Up to 97% increase in pressure head, 18% increase in hemolysis | Yes: hydraulics, hemolysis |
| | Wu et al. (2021)[9] | Manual comparison of the initial design and 3 variations | Centrifugal, maglev | Impeller blade inlet and outlet angles & blade thickness, radius of the leading edge of splitter blades | Turbulence intensity, secondary flows | Improved hydraulic performance, reduced turbulent intensities, reduced hemolysis | Yes: hydraulic performance |
| | Goodin et al. (2024)[10] | Manual comparison of the initial design and 3 variations | Mixed flow, axial contact bearing | Design of impeller blades; housing shape; number and design of diffuser vanes | Hydraulic performance, flow patterns | Improved hydraulic performance and reduced flow separation within the impeller and diffuser regions | Yes: hydraulic performance |
| **Systematic variation of design parameters** | Antaki et al. (2010)[11] | (Multi-disciplinary system-level optimization using reduced-order algebraic models for the fluid path, motor, suspension, heat transfer, and rotordynamics.) Initial pump flow path design using mean line analysis, followed by iterative CFD-informed evolution. | Axial flow, maglev | Thickness of the blades, radii of the impeller tip and root, wrap angle of the impeller blade, angle of the trailing edge, etc. | CFD: Hydraulic efficiency, hemolysis, thrombosis indices (stagnant, recirculating, or disturbed flow) | The CFD-informed evolution achieved satisfactory compromise between efficiency and hemolysis; the flow field was absent of major recirculation or stagnation | Yes: hydraulic performance, in vivo validation |
| | Wiegmann et al. (2018)[12] | Systematic variation of 3 design parameters generating 8 discreet designs | Centrifugal, bearing system unspecified | Impeller number of blades, clearance gap size, open/closed shroud | Flow rate, efficiency, hemolysis index, volume of fluid above shear stress threshold for hemolysis, thrombosis, and von Willebrand Factor cleavage. | Correlations & associations offered for studied design variables and performance metrics. Notably, hemolysis index showed a negative correlation with hydraulic efficiency | Yes: hydraulic performance |
| | Li et al. (2022)[13] | Systematic variation of 5 design parameters | Centrifugal with secondary flow passages, peg-top one-point bearing | Impeller number of blades, blade wrap angle, blade thickness, and splitter length & position | Pressure head, hemolysis, and platelet activation (stress accumulation) | Revealed the impact of design variables on the performance, as well as relationship between hemocompatibility, hydraulic performance, and flow characteristics | Yes: hydraulic performance |

| | Author | Method | Pump type | Design variables | Objective | Result | Experimental validation |
|---|---|---|---|---|---|---|---|
| **Automated optimization** | Burgreen et al. (1998)[14] | A first-order gradient based optimization algorithm | Axial flow, mechanical bearings | Outer housing shape of the outlet stator region | Maximizing area-weighted average of near-surface velocity magnitude, subject to the constraint of positive pressure rise | Improved surface washing while maintaining positive pressure rise | No |
| | Zhu et al. (2010)[15] | Nondominated sorting genetic algorithm, NSGA-II, generating 1637 design variants | Axial flow, mechanical bearings | Diffusor blade shape described by Bezier spline with 14 variables | Maximizing pressure head, minimizing backflow index | Set of Pareto-optimal designs maximizing pressure head and minimizing backflow index | No |
| | Yu et al. (2016)[16] | Evolutionary algorithm called "Evolution strategy" with 220 evaluated designs | Axial flow, bearing system unspecified | 7 design variables describing the shape of rotor blades, inlet and outlet guide vanes, shaft diameter, specific speed, etc. | Maximizing hydraulic efficiency | 25% relative improvement in hydraulic efficiency from initial design | No |
| | Mozafari et al. (2017)[17] | Initial design using similitude based on specific speed and specific diameter. Optimization using response surface based on Design of Experiments | Centrifugal, bearing system unspecified | Impeller number of blades, outlet angle, outlet width | Hydraulic efficiency, hemolysis | Relationships revealed between input variables and studied performance metrics. Optimal number of blades and geometric parameters offered for meeting desired pressure rise while limiting hemolysis | Yes: hydraulic performance |
| | Ghadimi et al. (2019)[18] | Multi-objective genetic algorithm assisted by an artificial neural network metamodel | Centrifugal, maglev | 11 design parameters describing the impeller blade and volute geometry | Maximizing hydraulic efficiency, minimizing hemolysis | 3% improvement in hydraulic efficiency and 12% reduction in hemolysis compared to the base design at the same OP | No |

| Study | Method | Pump type | Design parameters | Objectives | Results | Experimental validation |
|---|---|---|---|---|---|---|
| Nissim et al. (2023)[19] | Optimization using Genetic Algorithm. Surrogate models: Multi-linear regression, Gaussian Process Regression, Bayesian Regularized Artificial Neural Network. | Axial flow, maglev fore bearing, unspecified aft bearing | 5 design parameters controlling the shape of the impeller and diffuser blades | Maximizing hydraulic efficiency | 5.51% increase in efficiency at design point (a 20.9% performance increase) | Yes: hydraulic analysis |
| Escher et al. (2024)[20] | Multi-objective, multi-physics (hydraulic + electromagnetic) optimization using Pareto analysis | Cavo-pulmonary assist device for Fontan patients w/ double inflow and single outflow design, mechanical bearings | Gap width of the secondary flow channels, outer and inner rotor diameters, and stator height | Minimizing external dimensions of the pump and motor losses, minimizing/maximizing multiple hemocompatibility metrics | Pareto front containing 21 designs was obtained, revealing the trade-off between pump size, motor performance, and hemocompatibility. E.g. external volume could be reduced by 12.8% and hemocompatibility score improved by 10.1%, at the cost of 20% increase in motor losses. | No |
| Zhussupbekov et al. (2025) – *present study* | Multi-objective two-stage optimization using Sobol (exploration) and T-search (exploitation) algorithms | Axial flow, maglev | Newly added front diffuser stage with variable number of blades and blade shape controlled by 4 parameters | Maximizing pressure head, minimizing hemolysis | Best candidate was selected from Pareto-optimal set of diffuser designs balancing pressure recovery and hemolysis. Pump with added front diffuser achieved the same operating point (160±5 mmHg at 1.5 L/min) at a reduced speed of 14,000 RPM, with higher efficiency (32.5% vs 26.3%) and 31% lower computed hemolysis compared to the baseline pump at 16,000 RPM. | Yes: hydraulic performance |

**Supplementary Table S2.** Design input variables and their constraints used in Exploration 1 and 2.

| Input parameter | Exploration 1 | | Exploration 2 | |
|---|---|---|---|---|
| | Lower constraint | Upper constraint | Lower constraint | Upper constraint |
| Number of blades | 2 | 5 | 2 | 4 |
| Axial length (mm) | 6 | 18 | 8 | 19 |
| Wrap angle to axial length ratio (degree/mm) | 5 | 30 | For 2 blades: 10<br>For 3 blades: 8<br>For 4 blades: 8 | For 2 blades: 25<br>For 3 blades: 23<br>For 4 blades: 18 |
| Fullness scaling factor | 0.9 | 1.2 | 1.0 (fixed) | |
| Trailing edge angle | 70° | 90° | 85° (fixed) | |
| Leading edge angle | 10.26° (fixed) | | 10.26° (fixed) | |

**Supplementary Table S3.** Volume and percentage of the pump fluid path with shear rate exceeding $100\ s^{-1}$, justifying the assumption of constant viscosity used in the simulations.

| Pump version | Speed (RPM) | Q (LPM) | Volume of fluid path ($mm^3$) | Volume of fluid path $> 100 s^{-1}$ ($mm^3$) | % volume $> 100 s^{-1}$ |
|---|---|---|---|---|---|
| PF5 baseline | 14,000 | 1.5 | 2614.78 | 2608.29 | 99.75 |
| | 16,000 | 1.5 | 2614.78 | 2607.70 | 99.72 |
| PF5 w/ front stator | 14,000 | 1.5 | 2571.30 | 2545.33 | 98.90 |
| | 16,000 | 1.5 | 2571.30 | 2542.08 | 98.86 |

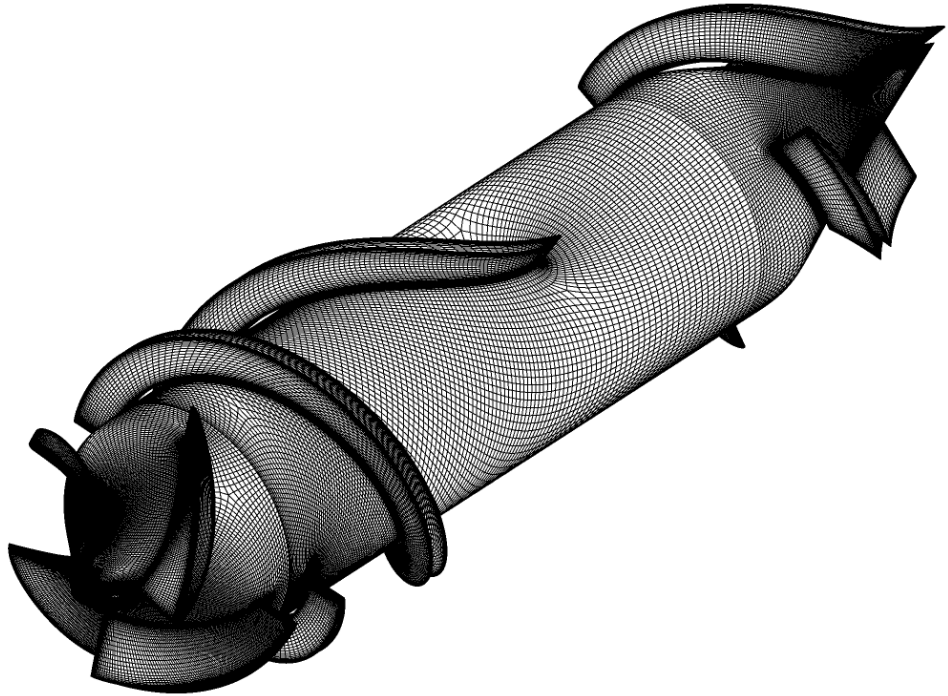

**Supplementary Figure S1.** Computational mesh used for CFD verification of the optimized front diffuser design. Multiblock structured mesh was created by Advanced Design Optimization, LLC, Irvine, CA, USA.

**Statistical Analysis**

Statistical relationships between design parameters and performance metrics were evaluated using multiple regression approaches appropriate for each variable type. For categorical variables (number of blades), one-way analysis of variance (ANOVA) was performed to test for group differences, with effect size quantified using eta-squared ($\eta^2$). For continuous variables (axial length, wrap angle to axial length ratio, trailing edge angle, fullness scaling factor), both linear and second-order polynomial regression models were fitted, and the model providing superior goodness-of-fit ($R^2$) was selected when the improvement exceeded 0.05.

Statistical significance was assessed at $\alpha = 0.05$, with relationships classified as strong ($R^2 > 0.3$), moderate ($R^2 = 0.1-0.3$), or weak ($R^2 < 0.1$). The coefficient of variation (CV) was calculated as the ratio of standard deviation to mean, expressed as a percentage, to quantify design space variance within and across parameter subsets.

For Exploration 1, analysis encompassed the full dataset (n = 210), while Exploration 2 analysis was performed separately for each blade configuration (n = 64 each for 2-, 3-, and 4-blade designs). Cross-blade comparisons utilized ANOVA with post-hoc linear trend analysis to quantify the effect of blade number on objective functions. All statistical analyses were conducted using Python with SciPy and scikit-learn libraries.

**Supplementary Table S4.** Exploration 1: Statistical assessment of parameter-performance relationships

| Parameter | Pressure Recovery | Primary Relationship | Action for Exploration 2 |
|---|---|---|---|
| Number of blades | $R^2 = 0.10$, $p < 0.001$ | Categorical, moderate effect | Adjust range |
| Axial length | $R^2 = 0.08$, $p < 0.001$ | Weak but significant, linear | Adjust range |
| Wrap angle to axial length ratio | $R^2 = 0.35$, $p < 0.001$ | Strong parabolic relationship | Adjust range |
| TE angle | $R^2 = 0.00$, $p = 0.323$ | No significant correlation | Eliminate (fixed value) |
| Fullness scaling factor | $R^2 = 0.00$, $p = 0.326$ | No significant correlation | Eliminate (fixed value) |

**Supplementary Table S5.** Exploration 2: Pressure and hemolysis performance by blade number configuration

| Number of blades (n = 64) | Pressure recovery | | Hemolysis, RHI | |
|---|---|---|---|---|
| | Mean, Std Dev (mmHg) | CV (%) | Mean, Std Dev (mmHg) | CV (%) |
| 2-blade designs | 34.26 ± 3.98 | 11.6 | 1.66 ± 0.07 | 4.3 |
| 3-blade designs | 30.67 ± 5.76 | 18.8 | 1.78 ± 0.11 | 5.9 |
| 4-blade designs | 28.15 ± 5.01 | 17.8 | 1.88 ± 0.14 | 7.4 |

**Supplementary Table S6**. Exploration 2: Effect of blade number on objective functions

| | -ΔP by Blade Number | Hemolysis, RHI, by Blade Number |
|---|---|---|
| Mean (n=64) | 2 blades: Mean = -34.26 mmHg<br>3 blades: Mean = -30.67 mmHg<br>4 blades: Mean = -28.15 mmHg | 2 blades: Mean = 1.66<br>3 blades: Mean = 1.78<br>4 blades: Mean = 1.88 |
| ANOVA Results | F-statistic: 24.428<br>p-value: 3.66e-10<br>Effect size ($\eta^2$): 0.205<br>Result: $p < 0.001$ | F-statistic: 62.223<br>p-value: 1.73e-21<br>Effect size ($\eta^2$): 0.397<br>Result: $p < 0.001$ |
| Linear Trend | $R^2$: 0.203<br>Slope: 3.06 per additional blade<br>p-value: 5.22e-11 | $R^2$: 0.396<br>Slope: 0.11 per additional blade<br>p-value: 1.36e-22 |

**Supplementary Table S7.** Exploration 2: Statistical assessment of parameter-performance relationships

| Parameter | Objective | 2-blade | 3-blade | 4-blade |
|---|---|---|---|---|
| Axial length | Pressure recovery | $R^2=0.32$, $p<0.001$ (Linear) | $R^2=0.15$, $p=0.001$ (Linear) | $R^2=0.28$, $p<0.001$ (Linear) |
| | Hemolysis | $R^2=0.52$, $p<0.001$ (Linear) | $R^2=0.75$, $p<0.001$ (Linear) | $R^2=0.76$, $p<0.001$ (Linear) |
| Wrap angle to axial length ratio | Pressure recovery | $R^2=0.28$, $p<0.001$ (Parab.) | $R^2=0.47$, $p<0.001$ (Parab.) | $R^2=0.34$, $p<0.001$ (Parab.) |
| | Hemolysis | $R^2=0.34$, $p<0.001$ (Linear) | $R^2=0.13$, $p=0.010$ (Parab.) | $R^2=0.10$, $p=0.012$ (Linear) |

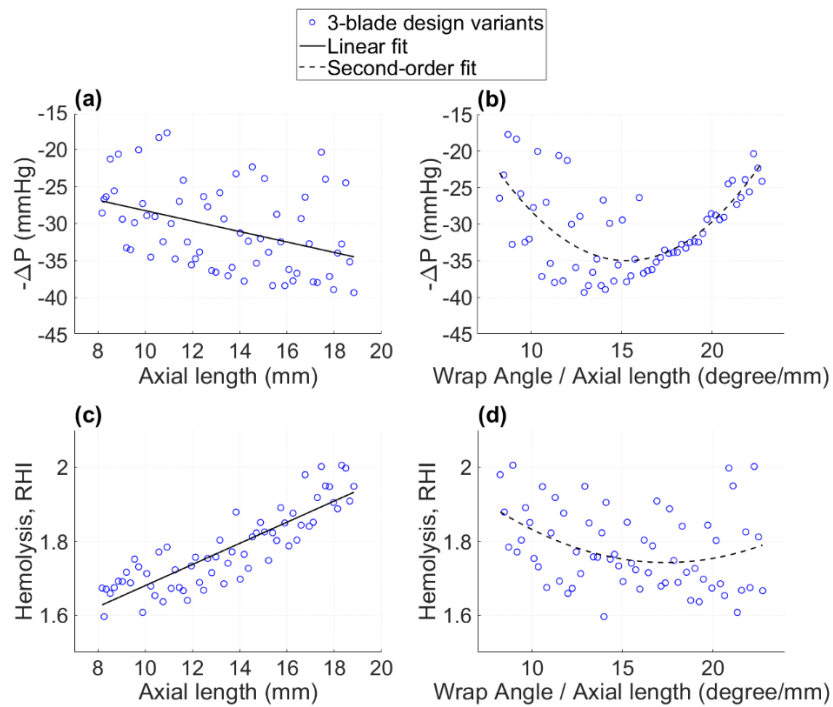

**Supplementary Figure S2.** Results of Exploration 2 for the 3-blade configuration. (a)-(b) Pressure recovery, $-\Delta P$, plotted against input variables. (c)-(d) Hemolysis, expressed as RHI, plotted against the input variables.

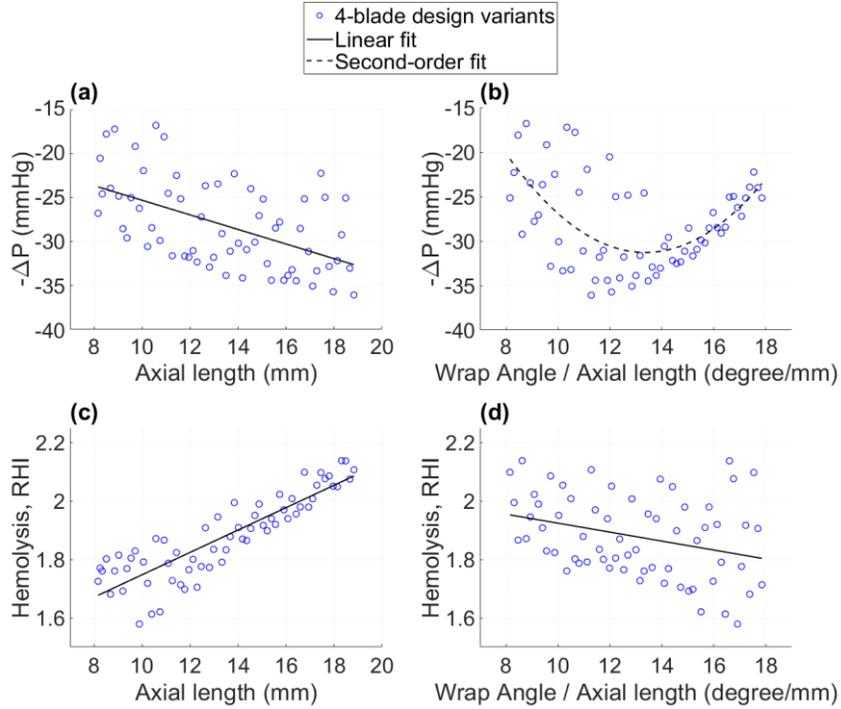

**Supplementary Figure S3.** Results of Exploration 2 for the 4-blade configuration. (a)-(b) Pressure recovery, -ΔP, plotted against input variables. (c)-(d) Hemolysis, expressed as RHI, plotted against the input variables.

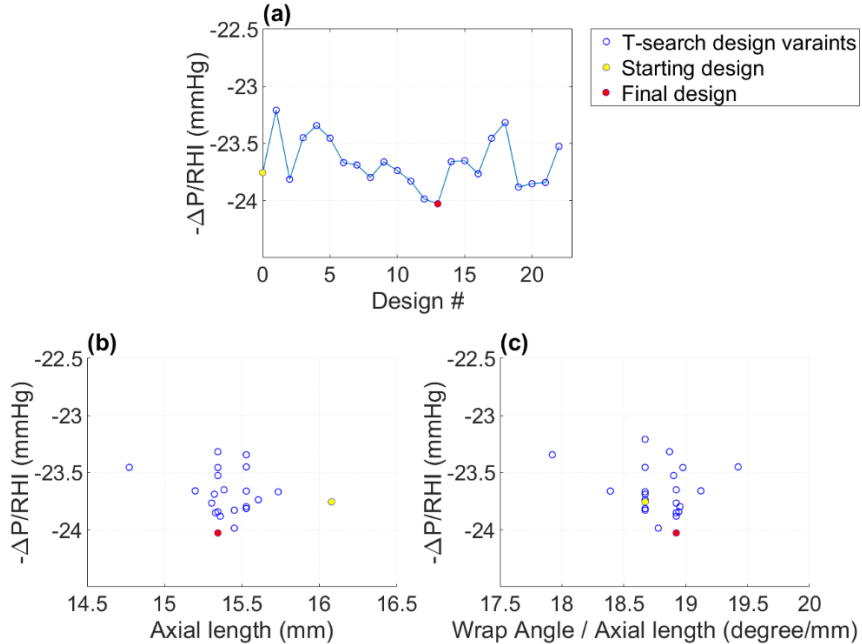

**Supplementary Figure S4.** In the final stage of optimization, the best candidate selected from the Exploration stage was further optimized using the T-search algorithm. (a) The selected design was optimized to improve the -ΔP/RHI ratio. The plot depicts the evolution of the objective function from the initial design in yellow to the optimized design in red. (b)-(c) The plots of the objective function against the input variables illustrate the search process conducted by the T-search algorithm within the design space.

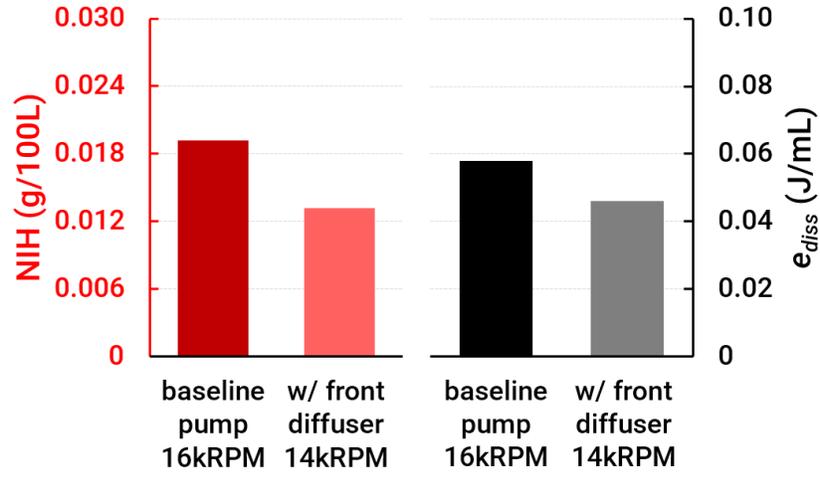

**Supplementary Figure S5.** CFD-predicted NIH and hydraulic energy dissipation, $e_{diss} = \frac{P_{loss}}{Q} = \frac{T \cdot \omega - H \cdot Q}{Q}$, of the baseline and modified pumps at 1.5 L/min and 160±5 mmHg.